\documentclass{INTERSPEECH2023}
\usepackage[symbol]{footmisc}
\usepackage{siunitx}
\usepackage{tablefootnote, stfloats, multirow}

\def\y{{\mathbf y}}
\def\w{{\mathbf w}}
\def\z{{\mathbf z}}
\def\c{{\mathbf c}}
\def\x{{\mathbf x}}
\def\t{{\mathbf t}}
\def\h{{\mathbf h}}

\interspeechcameraready

\title{FastFit: Towards Real-Time Iterative Neural Vocoder\\ by Replacing U-Net Encoder With Multiple STFTs}
\name{Won Jang, Dan Lim, Heayoung Park}
\address{
  Kakao Enterprise Corporation, Republic of Korea}
\email{\{taylor.martin, satoshi.2020, abigail.p\}@kakaoenterprise.com}

\begin{document}

\maketitle


\begin{abstract}
This paper presents FastFit, a novel neural vocoder architecture that replaces the U-Net encoder with multiple short-time Fourier transforms (STFTs) to achieve faster generation rates without sacrificing sample quality. We replaced each encoder block with an STFT, with parameters equal to the temporal resolution of each decoder block, leading to the skip connection. FastFit reduces the number of parameters and the generation time of the model by almost half while maintaining high fidelity. Through objective and subjective evaluations, we demonstrated that the proposed model achieves nearly twice the generation speed of baseline iteration-based vocoders while maintaining high sound quality. We further showed that FastFit produces sound qualities similar to those of other baselines in text-to-speech evaluation scenarios, including multi-speaker and zero-shot text-to-speech.
\end{abstract}
\noindent\textbf{Index Terms}: Neural vocoder, text-to-speech, U-Net, short-time Fourier transform

\section{Introduction}
Neural vocoders generate speech that conforms to the given input conditions by modeling short- and long-term dependencies. Owing to these features, these architectures have been applied\cite{van2016wavenet, kumar2019melgan}, wholly or partially, to various applications\cite{stoller2018wave, lee2022direct, zhang2022visinger} that output speech and audio as well as text-to-speech applications\cite{shen2018natural, lim2020jdi, ren2020fastspeech, kim2021conditional}. Moreover, the use of generative adversarial networks (GANs)\cite{goodfellow2014generative} for neural waveform generation has further improved neural vocoders\cite{kumar2019melgan, yamamoto2020parallel, kong2020hifi, jang2021univnet, kaneko2022istftnet, lee2022bigvgan}. However, according to recent text-to-speech studies, some vocoders require additional training (i.e., fine-tuning) using pairs of ground-truth waveforms and model-predicted features to adapt to low-quality audio features generated by an acoustic model\cite{lee2022direct, kong2020hifi, jang2021univnet, kaneko2022istftnet}.

Recent research has shown that image generation models utilizing denoising diffusion probabilistic models (DDPMs)\cite{ho2020denoising} outperform traditional GAN-based models\cite{dhariwal2021diffusion}. Several studies have successfully applied DDPM to neural vocoders, with some reporting superior performance over conventional models\cite{chen2021wavegrad, kong2021diffwave, huang2022fastdiff, koizumi2022specgrad}. However, the trade-off between generation speed and quality owing to the need for repeated denoising is considered a barrier to the commercialization of these models. Subsequent studies have attempted to overcome this by maintaining robust performance with fewer iterations\cite{huang2022fastdiff, chen2022infergrad, koizumi2022wavefit}.

The symmetric architecture of U-Net\cite{ronneberger2015unet} has made it an attractive choice for iteration-style models. To use an existing GAN-based vocoder as a decoder, some studies have added an encoder connected with skip connections\cite{chen2021wavegrad, huang2022fastdiff}. However, this doubles the size of the model and slows the generation speed by approximately half.

To improve efficiency, we propose FastFit, a new architecture that replaces the encoders in U-Net with multiple short-time Fourier transforms (STFTs) to trade a small fidelity degradation for a high generation speed gain. Our work is inspired by the work of Kaneko \textit{et al.}~\cite{kaneko2022istftnet}, who replaced some of the blocks with an inverse STFT. We extended the GAN-based vocoder proposed by Jang \textit{et al.}~\cite{jang2021univnet} to U-Net and replaced each encoder block with an STFT with parameters corresponding to the shape of its skip connection. This modification of the model preserves the advantages of skip connection in U-Net, while expecting more efficient intermediate feature encoding of raw waveforms because the computational cost of STFT is less than that of the neural encoder.

We applied the iteration-style principle proposed by Koizumi \textit{et al.}~\cite{koizumi2022wavefit} to the proposed architecture. To compare the performance of FastFit, we used iteration-based vocoders, which reported fast generation speeds as baselines and conducted objective and subjective evaluations. The results showed that FastFit achieved twice the generation speed with statistically similar speech quality despite having half the baseline parameters. Further, we conducted experiments by applying each vocoder to multi-speaker and zero-shot text-to-speech without fine-tuning, and FastFit was found to be one of the best-performing models\footnote{Audio demo samples can be found at the following URL: \\ \url{https://kallavinka8045.github.io/is2023/}}.

\label{fig:proposed_model}
\begin{figure*}[t!]
\begin{minipage}[t!]{0.3\linewidth}
  \centering
  \centerline{\includegraphics[height=5.2cm]{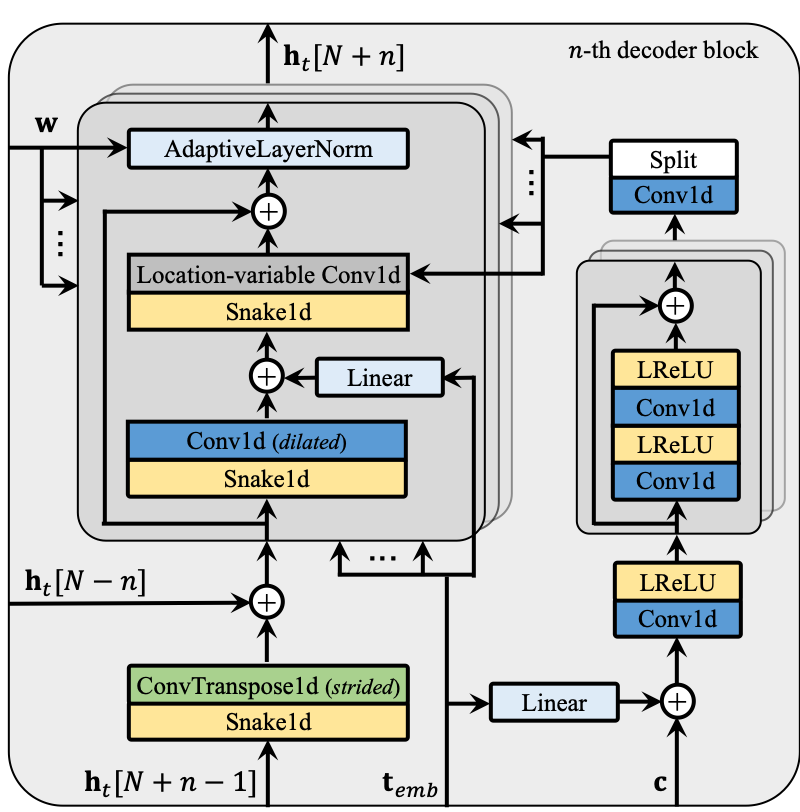}}
  \centerline{(a) $n$-th decoder block}\medskip
\end{minipage}
\begin{minipage}[t!]{0.41\linewidth}
  \centering
  \centerline{\includegraphics[height=6.3cm]{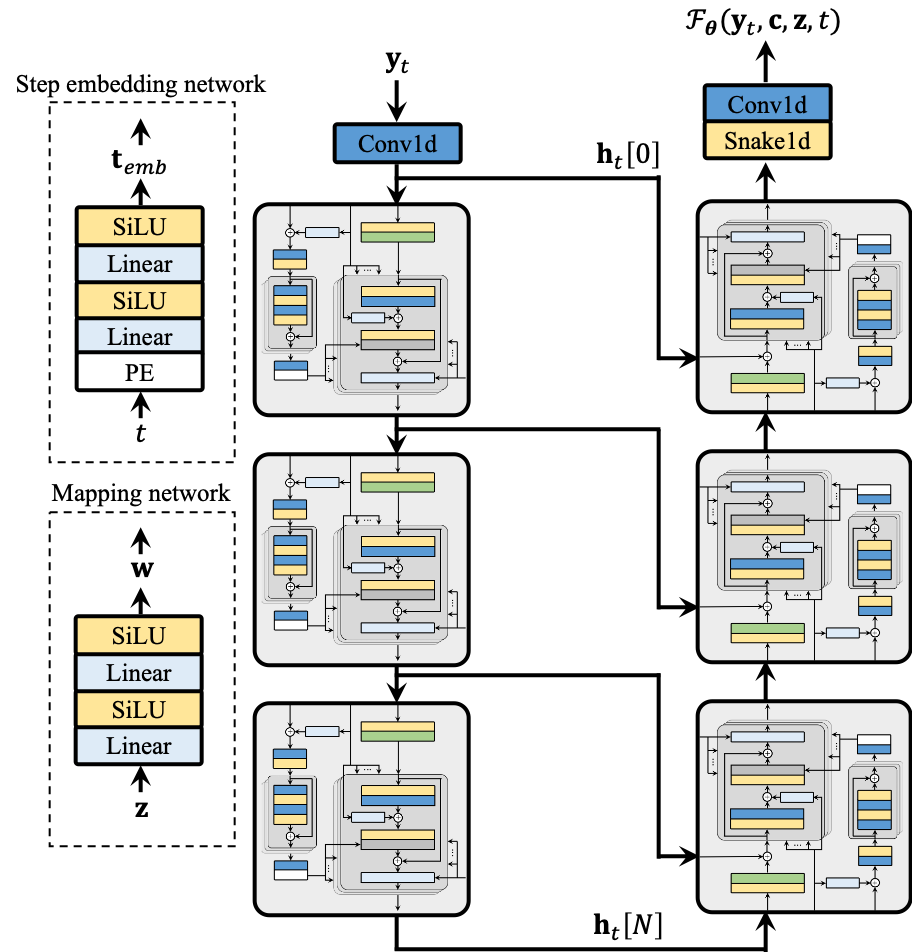}}
  \centerline{(b) FastFit (U-Net)}\medskip
\end{minipage}
\begin{minipage}[t!]{0.27\linewidth}
  \centering
  \centerline{\includegraphics[height=6.3cm]{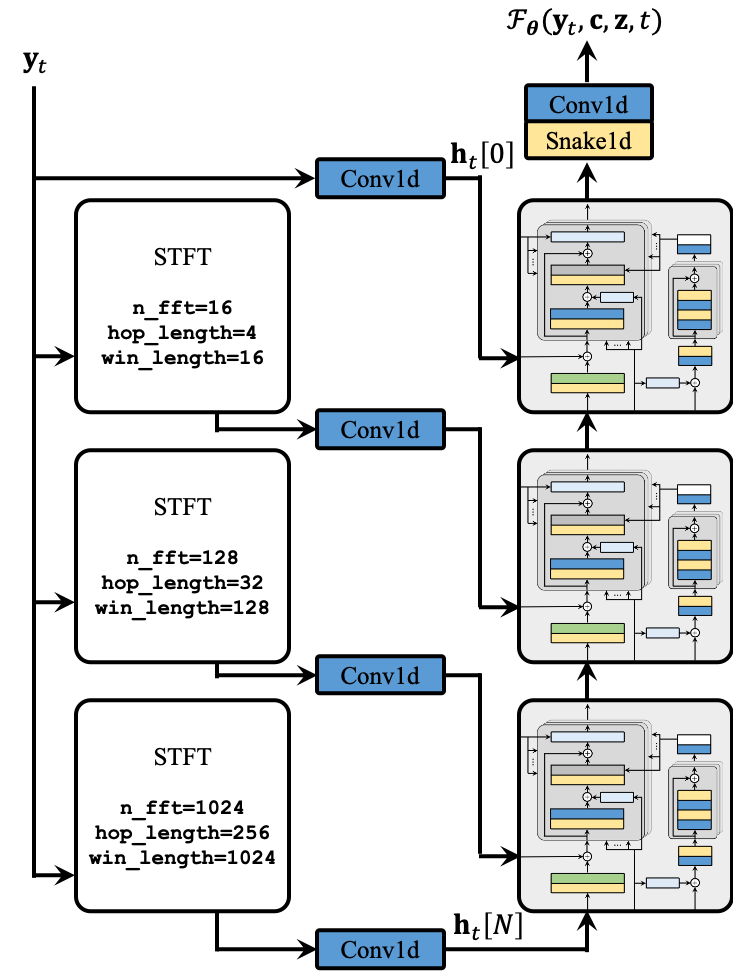}}
  \centerline{(c) FastFit (STFTs encoder)}\medskip
\end{minipage}
\caption{FastFit architecture. (a) The $n$-th decoder block. For example, the $1$-st decoder block computes $\h_t[N+1]$ from the output $\h_t[N]$ of the last $N$-th encoder block and the skip-connection $\h_t[N-1]$, the output of the $N-1$-th encoder. (b) FastFit based on U-Net. $\c$, $\w$, and $\t_{emb}$ are used as inputs to each block, but we omitted them in this figure for brevity. $\mathrm{PE}$ denotes a positional encoding operation\cite{vaswani2017attention}. (c) FastFit with multiple STFTs encoder, based on the proposed U-Net version. Each channel size of STFT is converted to fit the channel of the decoder block through each convolution layer.}
\end{figure*}

\section{Related work}

The proposed model is influenced by several improvements from previous studies on GAN-based vocoders. Parallel WaveGAN\cite{yamamoto2020parallel} applied multi-resolution STFT (MR-STFT) loss as an auxiliary loss to a vocoder to facilitate stable adversarial training. HiFi-GAN\cite{kong2020hifi} and UnivNet\cite{jang2021univnet} include a multi-period discriminator (MPD) and multi-resolution spectrogram discriminator (MRSD), respectively, which are discriminators that can observe real or generated waveforms with various patterns and resolutions. iSTFTNet\cite{kaneko2022istftnet} replaced the back part of the residual blocks of HiFi-GAN with an inverse short-time Fourier transform (iSTFT), trading off a small reduction in quality for higher synthesis speeds. Recently, BigVGAN\cite{lee2022bigvgan} succeeded in the adversarial training of large-scale generators with more than 100M parameters, achieving overall state-of-the-art fidelity, including out-of-distribution robustness.

Our work is based on the iteration-style vocoding principle proposed by WaveFit\cite{koizumi2022wavefit}. According to the fixed-point iteration theorem, if a mapping $\mathcal{T}$ has a fixed point $\x=\mathcal{T}(\x)$ and is firmly quasi-nonexpansive (as described in Section~17.2.2 in Yamada \textit{et al.}~\cite{yamada2011minimizing}), then the mapping point $\mathcal{T}(\y)$ of an arbitrary point $\y$ always has a smaller Euclidean distance from $\x$ than $\y$. $\mathcal{T}$ can be extended to the form of iterative denoising $\y_{t-1}=\mathcal{T}(\y_{t})$. If an arbitrary initial point $\y_T$ is iteratively refined at each $t$ from $T$ to $1$, then $\y_{t-1}$ always moves closer than $\y_t$ to the clean signal $\x$, which is the fixed point of $\mathcal{T}$. WaveFit proposed a denoising mapping and loss function for a vocoder that satisfies this property.

\section{Description of the proposed model}

The proposed model, FastFit, begins with an initial point $\y_T$. At each iteration step, $t=T,T-1,...,1$, denoising mapping is applied to $\y_t$ to obtain the denoised signal $\y_{t-1}$. A model $\mathcal{F}$ parameterized by $\theta$ was trained to predict the noise components of $\y_t$. $\mathcal{F}_{\theta}$ was conditioned on the log-mel-spectrogram $\c$, latent noise $\z\sim\mathcal{N}(0,\mathbf{I})$, and current step $t$ as $\mathcal{F}_{\theta}(\y_t,\c,\z,t)$. The objective of the vocoder is to make $\y_t$ at each iteration, including the final output $\y_0$, close to the target waveform, $\x$.

\subsection{Improving the architecture of the residual block}

Our U-Net model has $N$ encoder and decoder blocks with mapping and step embedding networks for the intermediate latent $\w$ and step embedding $\t_{emb}$, respectively, as shown in Figure~1(b). Each $n$-th decoder block computes $\h_t[N+n]$ with $\h_t[N+n-1]$, $\c$, $\w$, $\t_{emb}$, and $\h_t[N-n]$ as inputs, as shown in Figure~1(a). $\t_{emb}$ is conditioned to be broadcast and added to the following features after $\c$ and after each dilated convolution.

Each decoder block is based on the UnivNet\cite{jang2021univnet} generator with three main changes. First, we added an adaptive layer normalization (AdaLN) after each residual connection to inject noise $\z$ into the vocoder for improving the performance and training stability\cite{karras2019style}. Second, we applied the snake activation function\cite{lee2022bigvgan} to the model. This trainable activation function controls the output of each layer in the form of a periodic frequency and contributes to out-of-distribution robustness\cite{lee2022bigvgan}. Finally, the gated activation units were removed to improve the generation speed. Although this layer contributes to a slight improvement in quality according to UnivNet, it doubles the number of channels in the previous layer. The encoder block is like the decoder block, with a few differences: there is no skip connection, and the upsampling layer is replaced with downsampling using strided convolution layer.

\subsection{Replacing U-Net encoder with multiple STFTs}
To approach a real-time iterative vocoder application, we propose an intuitive methodology: replacing the U-Net encoder with multiple STFTs. As shown in Figure~1(c), we used a frame shift interval equal to the temporal resolution of each decoder block, leading to the skip connection as a parameter for each STFT. Inspired by iSTFTNet\cite{kaneko2022istftnet}, the number of points in the Fourier transform and Hann window length was set to four times the respective frame shift interval. To match the channel size, a convolution layer was placed between each encoder and decoder blocks. Because the computation speed of STFT is high compared to that of the neural encoder block, our proposed model could reduce the number of parameters by almost half, approximately doubling the generation speed. We expect limited degradation of speech quality using the methodology because skip connections are still used, which is the basis for the high performance of the U-Net architecture.

We conducted an ablation study to determine the optimal representations of STFT. Consequently, we chose the Cartesian form (concatenation of real and imaginary channels) as the STFT representation.

\subsection{Denoising mapping and training losses}

As mentioned in the previous section, our model is based on denoising mapping and the loss function proposed using WaveFit. The denoised signal $\y_{t-1}$ is then computed as follows:

\begin{equation}
\tilde{\y}_{t}=\y_{t}-\mathcal{F}_{\theta}(\y_t,\c,\z,t)
\end{equation}
\begin{equation}
\mathbf{y}_{t-1}=(P_\c/(P_{\tilde{\y}_{t}}+s))\tilde{\y}_{t}
\end{equation}

\noindent{}where $s=10^{-8}$ is a constant used to avoid numerical errors. The denoising mapping is defined by subtracting the noise component predicted by $\mathcal{F}_{\theta}$ from $\y_t$ to obtain $\tilde{\y}_{t}$ and adjusting the power of $\tilde{\y}_{t}$ to $P_\c$. $P_{\tilde{\y}_{t}}$ and $P_\c$ can be obtained by computing the power spectrograms of $\tilde{\y}_{t}$ and $\c$, respectively, and then taking the element-wise mean. Specifically, the power spectrogram of $\c$ can be obtained by multiplying $\c$ with the pseudoinverse of the mel-compression matrix and then squaring it. By scaling the power of the signal to a constant power of $\c$ at each step, the power of $\y_{t-1}$ can be kept constant until denoising is repeated for all $t$ and the final output $\y_{0}$ is obtained.

FastFit is adversarially trained with the least squares GAN (LSGAN)\cite{mao2017least} as the GAN loss and discriminators $D$, which are a combination of MPD (as described in Appendix~B.2 in Kim \textit{et al.}~\cite{kim2021conditional}) and MRSD\cite{jang2021univnet}. The overall losses $\mathcal{L}_{\mathrm{disc}}$ and $\mathcal{L}_{\mathrm{gen}}$ are defined as follows.

\begin{footnotesize}
\begin{equation}
\mathcal{L}_{\mathrm{disc}}=\frac{1}{TK}\sum_{t=0}^{T-1}\sum_{k=0}^{K-1}\Big{[}\mathbb{E}_{\x}[(D_{k}(\x)-1)^2]+\mathbb{E}_{\y_t}[D_{k}(\y_t)^2]\Big{]}
\end{equation}
\begin{align}
\mathcal{L}_{\mathrm{gen}}&=\frac{1}{T}\sum_{t=0}^{T-1}\bigg{[}\lambda_{\mathrm{aux}}{\mathcal{L}_{\mathrm{aux}}}(\y_t,\x)\nonumber\\&+\hspace{-1pt}\frac{1}{K}\hspace{-1pt}\sum_{k=0}^{K-1}\hspace{-1pt}\Big{[}\mathbb{E}_{\y_t}[(D_{k}(\mathbf{y}_{t})\hspace{-1pt}-\hspace{-1pt}1)^2]\hspace{-1pt}+\hspace{-1pt}\lambda_{\mathrm{fm}}{\mathcal{L}_{\mathrm{fm}}(D_{k};\y_t,\hspace{-1pt}\x)}\Big{]}\hspace{-1pt}\bigg{]}
\end{align}
\end{footnotesize}

\noindent{}where $K$ denotes the number of sub-discriminators. We used MR-STFT\cite{yamamoto2020parallel} as the auxiliary loss $\mathcal{L}_{\mathrm{aux}}$ and set $\lambda_{\mathrm{aux}}$ to 2.5. Additionally, we applied the scaled feature matching loss $\mathcal{L}_{\mathrm{fm}}$ proposed by Yang \textit{et al.}~\cite{yang2021ganspeech}; $\lambda_{\mathrm{fm}}=\lambda_{\mathrm{aux}}\mathcal{L}_{\mathrm{aux}}/\mathcal{L}_{\mathrm{fm}}$.

According to WaveFit, the initial point $\y_T$ is sampled using the noise generation algorithm of SpecGrad\cite{koizumi2022specgrad}, which is defined as follows:

\begin{equation}
\y_T=\mathbf{G}^{+}\mathbf{M}\mathbf{G}\boldsymbol{\epsilon}
\end{equation}

\noindent{}where $\boldsymbol{\epsilon}\sim\mathcal{N}(0,\mathbf{I})$, $\mathbf{G}$ and $\mathbf{G}^{+}$ denote STFT and iSTFT, respectively, and $\mathbf{M}$ denotes a filter computed from $\c$ for prior adaptation. SpecGrad estimated a cepstrum-based spectral envelope from the spectrogram obtained by multiplying $\c$ with the pseudoinverse of the mel-compression matrix (like the previous paragraph) and used it as $\mathbf{M}$. However, SpecGrad reported that using the spectrogram from $\c$ directly did not provide satisfactory results. In contrast, we experimentally verified that, unlike the case of the DDPM algorithm on which SpecGrad is based, using the spectrogram as $\mathbf{M}$ yields higher sound quality under our denoising mapping and architecture. The experimental design and results are described in detail in the following sections.

\section{Experiments}

\subsection{Data configurations and evaluation metrics}

We adopted LibriTTS\cite{zen2019libritts}, a multi-speaker English dataset with 24 kHz sampling rate waveforms, for the training and evaluation of the vocoder models. We used the “train-clean-360” dataset to train the models, with 5\% and 2\% of the dataset for validation and testing, respectively, with all speakers included in each of the three splits. For ground-truth mel-spectrogram evaluation (GT mel evaluation) including the ablation study, a “test-clean” dataset was prepared. The STFT parameters used to extract the 100-band, 0-12 kHz log-mel-spectrograms are 1024-point Fourier transform, 256 sample frame shift, and 1024 sample Hann window length. 

Two objective evaluation metrics, PESQ and MR-STFT, were used to evaluate the performance of each model. An open-source library\footnote{https://github.com/vBaiCai/python-pesq} was used to calculate the wideband PESQ, and the parameters required to calculate the MR-STFT metric were set to the same values as those in Yamamoto \textit{et al.}~\cite{yamamoto2020parallel}.

To clarify the comparison between the proposed model and baselines, a 5-point mean opinion score (MOS) was used for the TTS evaluation and a 7-point comparative MOS (CMOS) evaluation was used for GT mel evaluation. To collect 400 ratings for each evaluation item, we randomly sampled 20 speech samples and collected 20 ratings for each sample from 20 listeners located in the United States using Amazon Mechanical Turk. The loudness of all speech samples used was normalized to -23 LUFS. Other details of the subjective evaluations were based on Loizou's work\cite{loizou2011speech}.

\subsection{Model settings}

The proposed model uses the hyperparameters of each block that follow UnivNet-c32\cite{jang2021univnet}, with the number of dilated convolutions reduced to three, each with a dilation of \{1,3,9\}, to improve speed. The channel size of each convolution in the MRSD was set to 16. The dimensions of the latent noise $\z$ and the number of iterations $T$ were set to 100 and 3, respectively. The step embedding network used the structure proposed by Kong \textit{et al.}~\cite{kong2021diffwave}, and the mapping network used the same structure, but the channel size of each layer was set to 256. The minimum phase response based on the homomorphic filter method was used to calculate the filter $\mathbf{M}$. The same optimizer and learning rate as UnivNet were used to train FastFit and all the models for the ablation study up to 1M steps, with a batch size of 64. All other architectural details followed the settings of the studies on which they were based.

The performance of the proposed model was compared with three baselines: UnivNet, FastDiff, and WaveFit. These models are based on three main methodologies of vocoder research: GAN, DDPM, and fixed-point iteration. For UnivNet, we used the “c32” version of our implementation. FastDiff was implemented using the official repository\footnote{https://github.com/Rongjiehuang/FastDiff}, with $T=4$ as suggested by Huang \textit{et al.}~\cite{huang2022fastdiff}. We implemented WaveFit with $T=3$ following Koizumi \textit{et al.}~\cite{koizumi2022wavefit}, using an unofficial implementation\footnote{https://github.com/ivanvovk/WaveGrad} of WaveGrad with 15.8M parameters as the base model. The upsampling ratios were set to \{4,4,4,2,2\} to fit our experimental setting. To improve the training stability, we replaced the GAN loss with an LSGAN, which resulted in more stable training and a lower auxiliary loss. All the models were trained up to 1M steps using four NVIDIA V100 GPUs, and no additional fine-tuning was applied.

For the multi-speaker TTS evaluation, we trained the JDI-T\cite{lim2020jdi} acoustic model with the adapter-based multi-speaker methodology of Hsieh \textit{et al.}~\cite{hsieh2022adapter} using the LibriTTS train-clean-360 subset with 100 speakers. For the zero-shot TTS evaluation, we used an open-source zero-shot TTS program named TorToiSe\footnote{https://github.com/neonbjb/tortoise-tts}. The recordings of the LibriTTS “test-clean” subset with 10 speakers were input into the program with an “ultra-fast” offset to synthesize mel-spectrograms for evaluation. Sentences for all TTS evaluations were extracted from the utterances of speakers who were not included in the training. All speaker selection and data processing details for TTS evaluations followed Hsieh \textit{et al.}~'s approach.

\begin{table}[t]
\centering
\caption{The ablation study results.}
\renewcommand{\tabcolsep}{1.5mm}
\begin{footnotesize}
\begin{tabular}{l|ccc}
\toprule
Model                                    & PESQ$\uparrow$ & MR-STFT$\downarrow$ & CMOS$\uparrow$ \\
\midrule
Recordings                               & -              & -                   & 0.251 \\
\midrule
FastFit                                  & 3.712          & \textbf{0.866}               & - \\
FastFit (U-Net)                          & \textbf{3.754}          & 0.868               & 0.062 \\
\midrule
Without AdaLN                            & 3.411          & 0.974               & -0.168 \\
Without skip-connections                 & 3.449          & 0.936               & -0.199 \\
\midrule
$\y_T\sim\hspace{3pt}$Spectral envelope  & 3.422          & 0.969               & -0.175 \\
$\y_T\sim\hspace{3pt}$Griffin-Lim        & 3.685          & 0.872               & \textbf{0.069} \\
\midrule
Magnitude STFTs encoder                   & 3.677          & 0.875               & -0.031 \\
Polar STFTs encoder                   & \multicolumn{3}{c}{Failed to train} \\
Polar+Cartesian STFTs encoder                   & \multicolumn{3}{c}{Failed to train} \\
\bottomrule
\end{tabular}
\end{footnotesize}
\end{table}

\section{Results}

\subsection{Ablation studies}
We conducted ablation studies to evaluate the proposed improvements under the GT mel conditions. According to the results presented in Table~1, FastFit achieved the non-significant objective and subjective metrics compared with U-Net with a neural encoder. We observed significant destabilization of the training when the model was trained without AdaLN, resulting in worse metrics. To demonstrate that the proposed methodology maintains quality by maintaining skip connections, we removed all skip connections and connected only one STFT to the first decoder block, which resulted in relatively poor metrics.

Moreover, we have made several attempts to define an effective initial point $\y_T$. Our experiments computing $\y_T$ from the spectral envelope performed worse than the proposed method of computing $\y_T$ directly from the spectrogram from $\c$. We also attempted to convert the spectrogram to a waveform with 32 Griffin-Lim iterations and use it as $\y_T$. This did not produce significant differences in metrics but resulted in a slight reduction in generation speed, so we did not adopt it.

We were intrigued by Webber \textit{et al.}~\cite{webber2022autovocoder}'s work with different representations of STFT output and tested to find the optimal one. The magnitude spectrogram, used as a representation, showed no significant metric difference from the Cartesian form of the proposed model. However, internal tests showed a decrease in quality in some samples; therefore, we did not adopt it. For the polar form, training collapsed early on, so using phase as a representation was not appropriate for the proposed model.

\begin{table*}[b]
\centering
\caption{Results of comparison with baseline models. “\textup{Speed}” indicates each model's generation speed relative to real time.}
\begin{footnotesize}
\begin{tabular}{l|cc|ccc|cc}
\toprule
                & \multicolumn{2}{c|}{Model complexity} & \multicolumn{3}{c|}{GT mel evaluation} & Multi-speaker TTS & Zero-shot TTS \\
\midrule
Model           & Params$\downarrow$ & Speed$\uparrow$ & PESQ$\uparrow$ & MR-STFT$\downarrow$  & CMOS$\uparrow$ & MOS$\uparrow$ & MOS$\uparrow$  \\
\midrule
UnivNet         & 14.86M & \textbf{$\times$314.49} & 3.705 & \textbf{0.853} & -0.295 & 3.46$\pm$0.06 & 3.84$\pm$0.09 \\
FastDiff        & 15.36M & $\times$52.35 & \textbf{3.786} & 1.385 & \textbf{0.078} & 3.32$\pm$0.07 & 3.68$\pm$0.10 \\
WaveFit         & 15.85M & $\times$43.10 & 3.639 & 0.921 & 0.072 & 3.70$\pm$0.08 & 3.83$\pm$0.09 \\
FastFit         & \textbf{6.81M} & $\times$101.40 & 3.712 & 0.866 & -     & 3.67$\pm$0.08 & 3.86$\pm$0.08 \\
FastFit (U-Net) & 12.94M & $\times$59.88 & 3.754 & 0.868 & 0.062 & \textbf{3.75$\pm$0.07} & \textbf{3.90$\pm$0.09} \\
\midrule
Recordings      & -      & -        & -     & -     & 0.251 & -             & -             \\
\bottomrule
\end{tabular}
\end{footnotesize}
\end{table*}

\subsection{Comparison with baselines}
To measure the speed of the evaluation models, we generated 6 second segments 20 times using an NVIDIA V100 GPU and measured the average time. As shown in Table~2, FastFit achieved approximately twice the synthesis speed of the iteration-based vocoders despite having approximately half the number of parameters, and none of the metrics scored significantly worse in terms of speech quality. Although UnivNet's synthesis speed was superior to other baselines, it performed poorly in the CMOS evaluation owing to the occasional blurring of the harmonic component of the GT mel-spectrogram. FastDiff performed best on PESQ but worst on MR-STFT, which calculates the numerical distances. This is because the model confused segments with noise-like spectral shapes, such as consonants, breaths, and high-frequency spectral bands, with noise and denoised them. WaveFit performed well overall but had the slowest generation speed.

\subsection{Application to text-to-speech synthesis}
We characterized the mel-spectrograms generated by the models for two TTS evaluation tasks and found that the mel-spectrograms for multi-speaker TTS had relatively blurry shapes because they were not generated by high-quality models, such as GAN or DDPM. In contrast, the mel-spectrograms for zero-shot TTS had relatively more realistic shapes because they used a DDPM-based acoustic model that produced high-quality output regardless of the synthesis rate. However, each vocoder was not fine-tuned using the predicted mel-spectrograms.

For multi-speaker TTS, FastFit produced better MOS scores than UnivNet and FastDiff, with a slight difference from WaveFit. For the zero-shot TTS, all models except FastDiff produced similar MOS. UnivNet was observed to produce noise artifacts in some segments of the multi-speaker TTS experiment, which may be responsible for its worse MOS. FastDiff recorded the worst MOS owing to the incorrect denoising of noise-like components and produced blurred harmonic components. The remaining models produced statistically similar MOS, with FastFit (U-Net) producing the best results, but the confidence intervals of the MOS overlapped.

\section{Conclusion}
By improving the architecture of an iteration-based neural vocoder, we could double the generation rate while maintaining a high fidelity. As the U-Net architecture is widely used in speech processing applications, we expect our simple yet effective idea of replacing the encoder with STFTs to be applied in a variety of speech-based research and applications in the future.

\section{Acknowledgements}
The authors would like to thank James Bekter for developing TorToiSe, an outstanding open-sourced TTS model. We also would like to thank Jaesam Yoon, Sunghee Jung, Gyeonghwan O and Bongwan Kim for providing insightful feedback.

\bibliographystyle{IEEEtran}
\bibliography{mybib}

\end{document}